\documentclass[amsmath,amssymbl]{revtex4}

\usepackage{graphicx}

\usepackage{bm}

\begin{document}

\title{Resonant Plasma Wave Growth and Monoenergetic Electron Beam Production using Collinear High-Intensity Ultrashort Laser Pulses}

\author{A.~G.~R.~Thomas$^1$}
\author{C.~D.~Murphy$^{1,2\dag}$}
\author{S.~P.~D.~Mangles$^1$}
\author{A.~E.~Dangor$^1$}
\author{P.~Foster$^2$}
\author{J.~G.~Gallagher$^3$}
\author{D.~A.~Jaroszynski$^3$}
\author{P.~A.~Norreys$^2$}
\author{R.~Viskup$^3$}
\author{K.~Krushelnick$^{1\ddag}$}
\author{Z.~Najmudin$^1$}

\affiliation{$^1$Blackett Laboratory, Imperial College London, SW7 2BZ, UK}
\affiliation{$^2$Central Laser Facility, Rutherford Appleton Laboratory, Oxon, OX11 0QX,  UK}
\affiliation{$^3$Department of Physics, University of Strathclyde, Glasgow, G4 0NG, UK}

\begin{abstract}
The resonant generation of relativistic plasma waves and plasma wave guiding by two co-propagating laser pulses has been studied. By proper timing between guiding and driver pulses, a resonant interaction occurs, which generates a high-amplitude plasma wave over a longer length than is possible with either of the laser pulses individually. The growth of the plasma wave is inferred by the measurement of monoenergetic electron beams with low divergence that are not measured by using either of the pulses individually. This scheme can be easily implemented, and allows more control of the interaction than is available to the single pulse scheme.
\end{abstract} 

\maketitle
The laser wakefield accelerator (LWFA) \cite{tajima} consists of a high-intensity laser pulse that generates an electron plasma wave with relativistic phase velocity, and oscillating at the electron plasma frequency  , $\omega_p = \sqrt{e^2n_e/m_e\epsilon_0}$, where $n_e$ is the local electron number density. Electrons that are not part of the relativistic fluid can be trapped in the wave, under certain conditions, and accelerated.

It is hoped that electron beams from LWFAs will eventually provide sources of relativistic electron beams and x-rays for applications, such as chemical and biological spectroscopy, medical imaging and radiation therapy. If viable, they are expected to revolutionize such sources, making them available at lower cost to small scale facilities such as universities or hospitals.

With advances in laser technologies, pulses of duration close to the plasma period ($\tau_p =2\pi /\omega_p)$ can now be generated directly. It was suggested that a high-amplitude wakefield could be generated by multiple suitably spaced short pulses \cite{rpw}; if subsequent laser pulses are timed to push the electrons in their direction of motion as they overshoot their equilibrium position, then the plasma wave amplitude can be grown resonantly. However, as pulse length reduction has been accompanied by an increase in the power of the short pulses, multiple pulses have been deemed to be unnecessary since it is now possible to grow large amplitude plasma waves with a single intense ultrashort laser pulse.

These high-amplitude waves can be a source for the trapped electrons by self-injection from the plasma itself. This occurs in three dimensions when the accelerating phase of the electric field reaches sufficient amplitude to prevent electrons from slipping into a decelerating phase of the wake \cite{gordy}. In recent experiments \cite{mangles,G_F}, it has been demonstrated that a single pulse injection process can produce electron beams of narrow energy spread and small divergence, but requires $>10\; TW$, $<40\;fs$ laser systems, which are by no means ubiquitous. In addition, since the propagation of a pulse---with $I\lambda^2>10^{19}\;\rm{Wcm^{-2}}$ required for self-injection---is dominated by modulational effects, a higher degree of complexity may be required to control the interaction.

Alternative means of injecting electrons by the influence of secondary laser beams have also been previously suggested \cite{umstadter, esarey, schroeder}. These include a scheme whereby counter-propagating laser pulses have been shown to control some of the electron beam properties \cite{faure2}. This demonstration has spurred further interest in optical injection in LWFAs.

In this paper we present the first evidence for production of mononenergetic electron beams by all optical injection in a dual collinear ultrashort laser pulse geometry  (i.e.~with pulses overlapping, parallel and traveling in the same direction). We note that though collinear pulses were used in \cite{gordon, Chen}, in those the resulting spectrum was not monoenergetic and the mechanism was different. The resonant growth of the plasma wave is experimentally inferred by the production of monoenergetic electron beams by the two laser pulses, when correctly spaced, as neither of the pulses on their own do this.

A laser pulse with a large focal spot ($f$-number) was used to generate a low amplitude plasma wave. This was then used as a guiding structure for a tightly focused (low $f$-number) pulse. Particle-in-cell simulations of the experiment show that the tightly focused pulse was guided  at a smaller spot-size than the guiding pulse, for much longer than its Rayleigh range if correctly phased within the wakefield \cite{sprangle:pra}. Electrons were produced when the low $f$-number pulse focuses, and are accelerated in the resonantly driven wake, resulting in monoenergetic beams.

\begin{figure}
\begin{center}
\includegraphics[width = 3.2in]{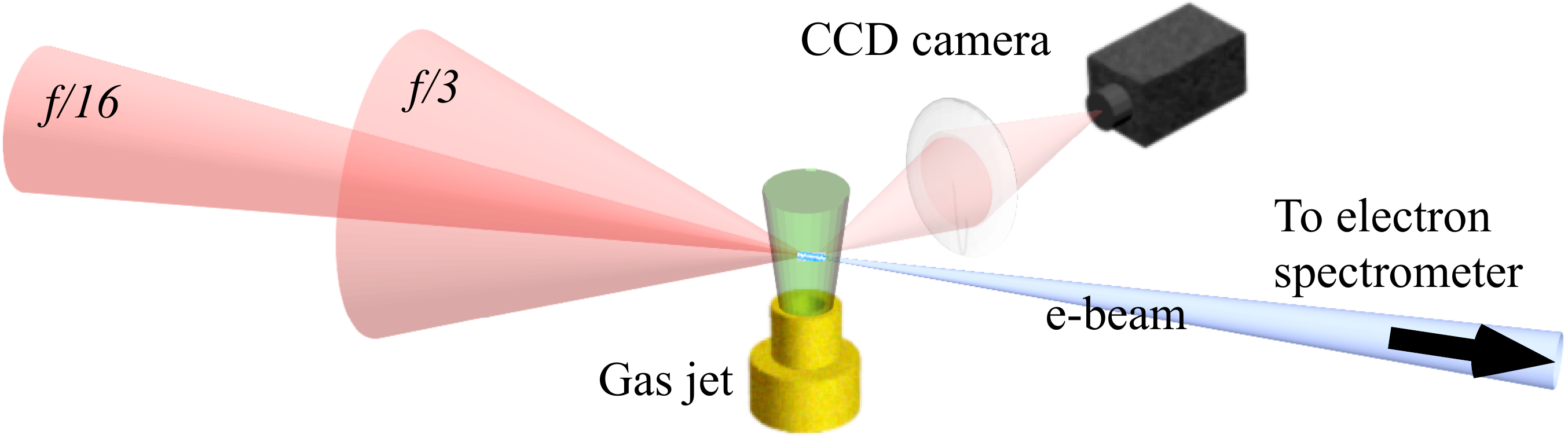}
\caption{(color online) Schematic of the experimental setup.}
\label{diag}
\end{center}
\end{figure}

The experiments were carried out on the 600 mJ arm of the $\tau_L=$ 40$\pm$5 fs full-width half-maximum (FWHM) Ti:Sapphire {\sc Astra} laser operating at wavelength $\lambda_0 = 800$ nm. A thin (5~mm) beam-splitter was used to produce two $E_{pulse}=300$~mJ pulses that were focused using $f/3$ and $f/16$ off-axis parabolic mirrors collinearly onto the edge of helium flow from a $2$ mm diameter supersonic gas nozzle (Fig.~\ref{diag}). On ionisation by the leading edge of the laser, a plasma of electron number density $ n_e=1\times 10^{19}\:\rm{cm}^{-3}$ was produced. This density was chosen so that the pulse with $f/16$ focusing did not produce a measurable electron signal, but so that electrons were detected using $f/3$ focusing.

The full width half maximum (FWHM) spot sizes for the pulses focused by the $f/3$ and $f/16$ optics were 5 $\rm{\mu m}$ and 25 $\rm{\mu m}$ respectively. This  resulted  in focused intensities of $1.9\times10^{19} \;\rm{Wcm^{-2}}$ and $0.8\times10^{18} \;\rm{Wcm^{-2}}$ corresponding to normalized vector potentials $a_{0}$ of 3 and 0.6. In both cases, the longer focal length  pulse passed through the beamsplitter, for which a $5$~fs increase in pulse length was measured. A motorized timing slide allowed control of the longitudinal spacing of the two pulses. The pulses could be overlapped to an accuracy of about one focal spot size (i.e.~$\sim25\, \mu$m), and were limited by the mechanical stability of the laser pointing at the time.

The timing slide position was measured to $\pm1\;\rm{\mu m}$  ($\pm3.3$ fs) accuracy. 
Accuracy was limited by the measured relative timing of the two pulses $\Delta t$, which was done by maximizing the plasma defocusing of one pulse by the other, at a fraction of a percent of full power. The error in this method was therefore of the order $40$ fs (i.e.~$\sim \tau_L$).
The energy spectrum of the accelerated electrons was obtained using a magnetic spectrometer with image plates as the detector, as in \cite{mangles}. 
The acceptance cone of the electron spectrometer was $f/200$. Light emitted from the interaction was reimaged orthogonally to the beam axis, to discern the propagation of the laser.

\begin{figure}
\begin{center}
\includegraphics[width = 3.2in]{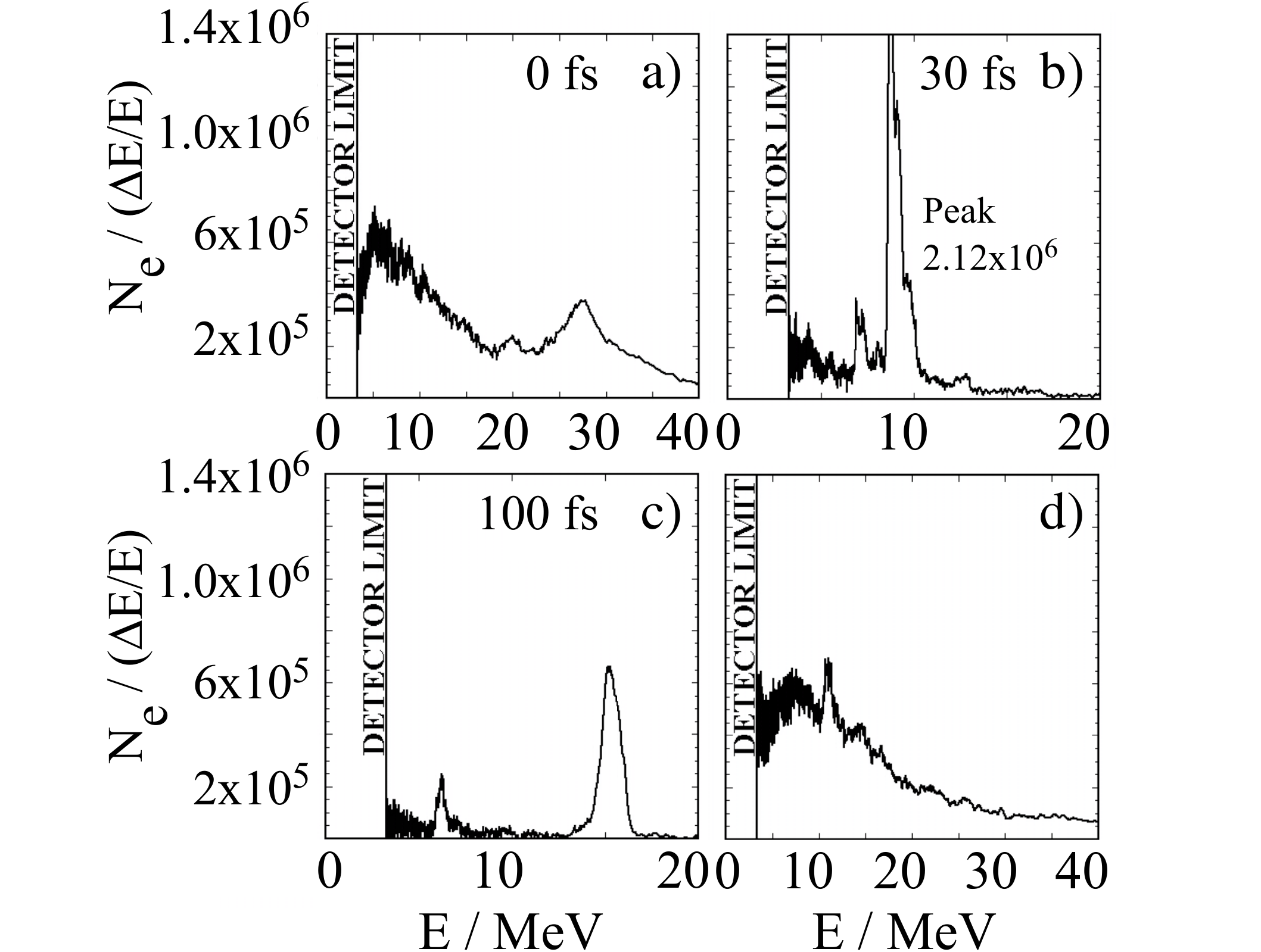}
\caption{(a-c) Quasi-monoenergetic electron spectra with highest charge shown for various timings between the drive and guiding laser pulses, compared with d) typical electron spectra with the drive pulse only. $n_e = 1 \times 10^{19}\;\rm{cm^{-3}}$ and  $E_{pulse}<300$ mJ.}
\label{Timespec}
\end{center}
\end{figure}

Using the pulse with $f/16$ focusing---henceforth referred to as the guiding pulse---alone resulted in no measurable electron signal above noise level at the chosen density. The pulse with $f/3$ focusing---henceforth referred to as the drive pulse---would consistently produce electron bunches of a few pC charge, but these were consistently best described by a single temperature (i.e.~one dimensional maxwellian {\it momentum} distribution) fit. 

Using both pulses, the spectra obtained were non-maxwellian with obvious structure (Fig.~\ref{Timespec}a-c). Indeed on many of the shots the maxwellian spectra that the drive pulse alone produced were almost completely suppressed, with the majority of recorded electrons in a single low energy spread beam. For example, the beam shown in Fig.~\ref{Timespec}c had an energy spread $\Delta E=0.6$~MeV (FWHM), which is the narrowest energy spread beam that we have yet measured from a self-injecting plasma accelerator. This monoenergetic beam was also very well collimated, with an angular divergence less than the opening angle of the  collimator, $2\times 10^{-5}$~steradians solid angle or $2.5$~mrad half-opening angle. This also indicates that the measured energy spread in this case is also determined by the actual size of the beam, and so was probably less than quoted.

\begin{figure}
	\begin{center}
	\includegraphics[width = 3.2in]{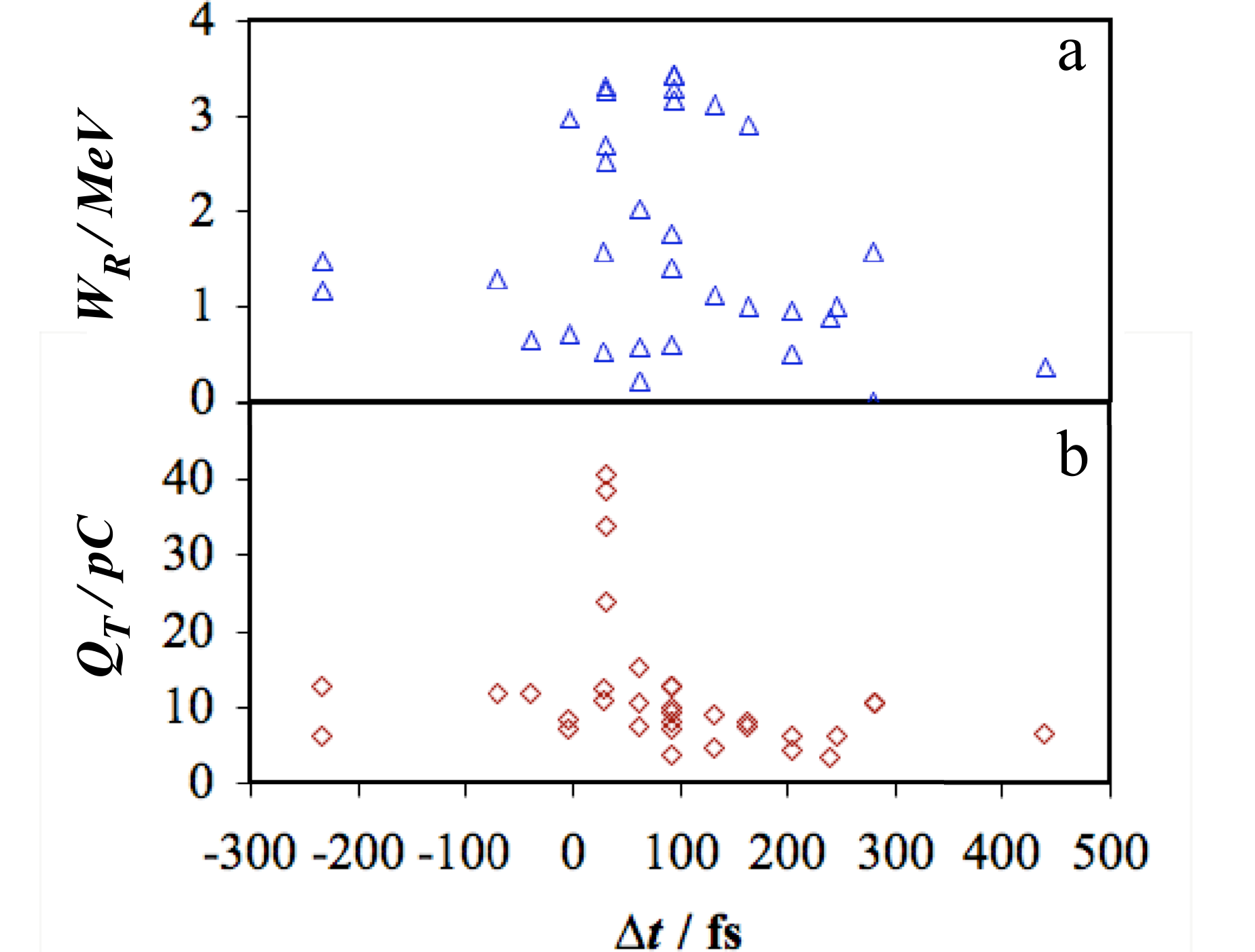}
	\caption{(color online) a) Relative mean energy $\overline{W}_R$ and b) total charge $Q_T$ as a function of the relative arrival time between the pulses for all shots.}
	\label{e-trel}
	\end{center}
	\end{figure}

The relative mean energies, $\overline{W}_R$, and total charge in the spectrum are plotted for all shots in the data set in Fig.~\ref{e-trel}, including non-monoenergetic spectra. The mean energy $\overline{W}= \int_0^\infty W\;dN(W)/ \int_0^\infty dN(W)$. $\overline{W}_R=\overline{W}-\min(\overline{W}_R)$. The total charge has a maximum for  $\Delta t = 30$ fs, which is likely to be the timing when the pulses were overlapped sufficiently to act as a single pulse with double the energy, since it is within the error by which $\Delta t = 0$ was set (Fig.~\ref{e-trel}b).

From Fig.~\ref{e-trel}a, it is evident that there is also an enhancement in the mean energy, but for a larger range of timings. These represent timings from when the pulses are overlapped, to when the guiding pulse leads the driver by a few plasma periods ($\approx 6$). This indicates the extent of the coherent plasma wave structure for this density. Note that due to the low intensity of the guiding pulse, a single wave-period is not expected as in \cite{pukhov}. However, the decay of the energy enhancement indicates the eventual damping of the coherent plasma wave due to, e.g. radial anharmonicity \cite{dawson}.

The oscillatory behavior of the weighted energy with increased timing appears to have a well defined frequency. However this is likely to be an artifact. Periodic behavior is expected due to the underlying plasma wave structure, but the frequency here is too low at $\omega\approx \omega_p/2$. However, the pulse length ($\sim 12\, \mu$m) was longer than $\lambda_p\sim 7.5\, \mu$m. Hence, it is likely that the periodic behavior is smeared out by the limiting pulse length. In addition, in our set-up the error in the relative timing was on the order of $\lambda_p$ due to variations in pulse length and positioning variation. This is reflected in the scatter of data points in Fig.~\ref{e-trel}. The oscillatory behavior is probably a combination of these errors and also the aliasing of an underlying frequency (i.e.~of the plasma wave) which is higher than that in the measurements. Nevertheless, it is worth emphasising that $\overline{W}_R$ can be enhanced but only when the driver follows within a time shorter than the decay time of the guide pulse wake. Yet over the same time-window, small variations in $\Delta t$ can also lead to a decrease in $\overline{W}_R$.

Another difference for the interaction with two beams is the images of side-scattered light. It has been demonstrated that this side-scatter radiation is primarily Raman side-scatter, and its length is an indication of the initial evolution phase of the high intensity laser plasma interaction \cite{thomas}. With the drive pulse alone, the extent of the emission is $\sim 10\; \rm{\mu m}$, which is comparable to the Rayleigh range of this pulse. Side-scatter of the guiding pulse by itself was not measurable above noise level. However, when the drive pulse overlaps the guiding pulse, an extended emission region on the order of 1~mm is observed, which is longer than the Rayleigh range of the guiding pulse. 

An extensive series of two dimensional particle-in-cell simulations were run using the code \textsc{osiris} \cite{osiris}. These were run under similar parameters to the experiments and for a large number of relative arrival times, $\Delta t_a$ between the pulses. Due to computational constraints, $\Delta t_a$ could not be longer than a few plasma periods, however the control of the timing within that range was obviously much higher than in the experiment.

The simulations clearly indicate the presence of plasma wave guiding \cite{sprangle:pra} of the tightly focused pulse. This counteracts the filamentary behavior of the tightly focused pulse and extends its propagation as a single filament. There is a strong phase dependence to this effect in these simulations arising as a result of the fine control of $\Delta t_a$. Plasma wave guiding
relies on the guided pulse being in a predominantly focusing refractive index structure. Since a plasma wave is periodically focusing and defocusing it is evident that for certain $\Delta t_a$ the pulse will be guided and for a $\pi/2$ phase shift the pulse will be defocused. 

\begin{figure}
\begin{center}
\includegraphics[width = 3.2in]{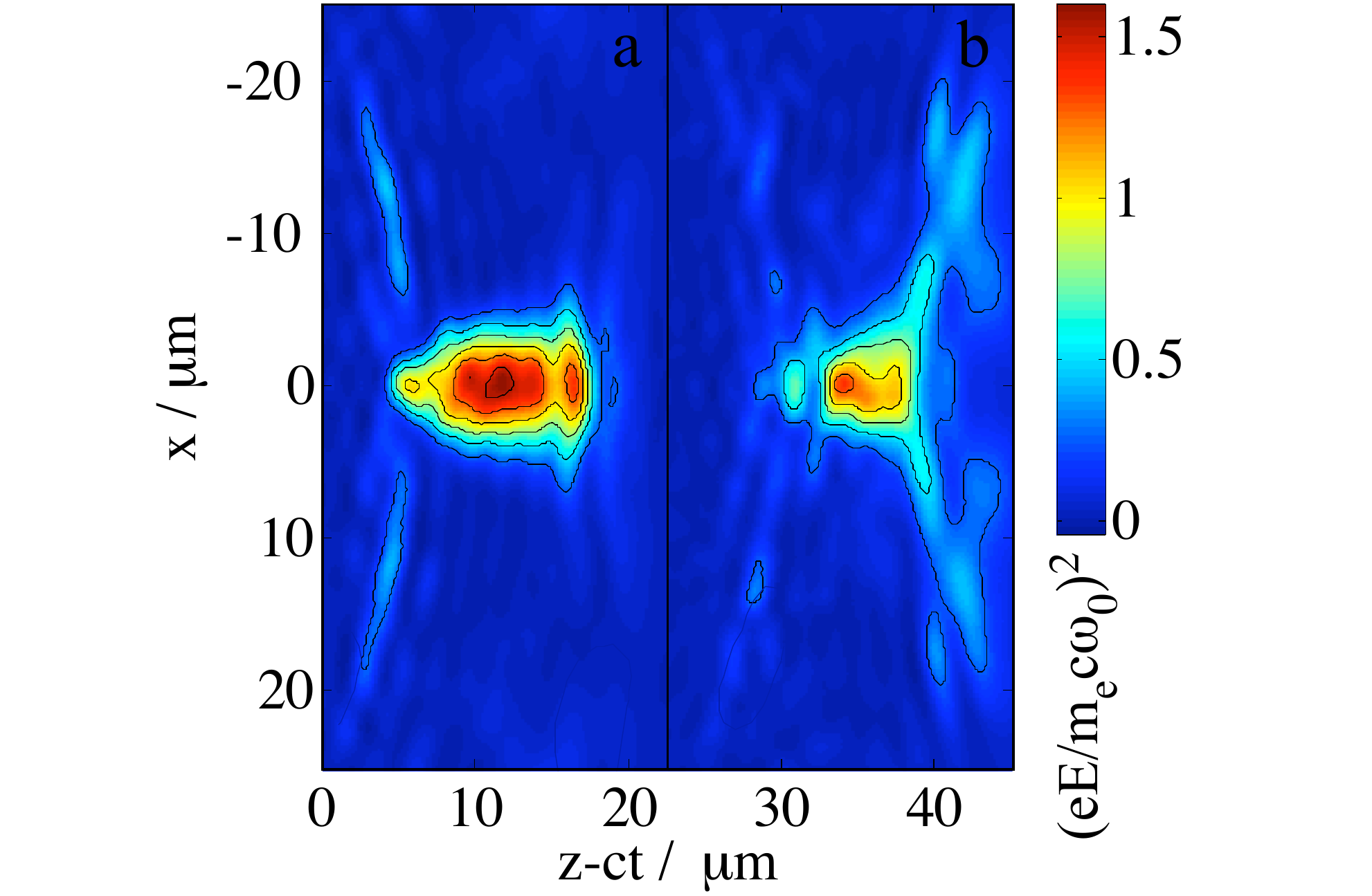}
\caption{(color online) The normalised intensity $e^2E_0^2/m^2c^2\omega_0^2$ of driver pulses having propagated $1500\mu m$ in a background density of $n_e=1\times 10^{19}$ with a separation of a) $21\mu m$ (well phased) and b) $18\mu m$ (not well phased). Contours are given at intervals of $0.25$.}
\label{pulses}
\end{center}
\end{figure}

For simulations when the drive pulse is in phase with the plasma wave created by the guiding pulse, it is focused by the density depression and its propagation is dictated by the guiding pulse. In addition, the two wakefields are resonant, and therefore the combination a larger amplitude wake than either pulse alone. When it is out of phase with the wavebucket, the density structure acts to defocus the pulse. This guiding and defocusing is shown in Fig~\ref{pulses}. In Fig~\ref{pulses}b the wake of the guiding pulse is out of phase with the drive pulse. The drive pulse is mainly defocused with only a small proportion of laser energy trapped in the density well. However, Fig~\ref{pulses}a shows the situation when the drive pulse phase is well matched within the guiding wake, and hence a large fraction of the pulse energy is trapped in a single stable filament.

\begin{figure}
\begin{center}
\includegraphics[width = 3.2in] {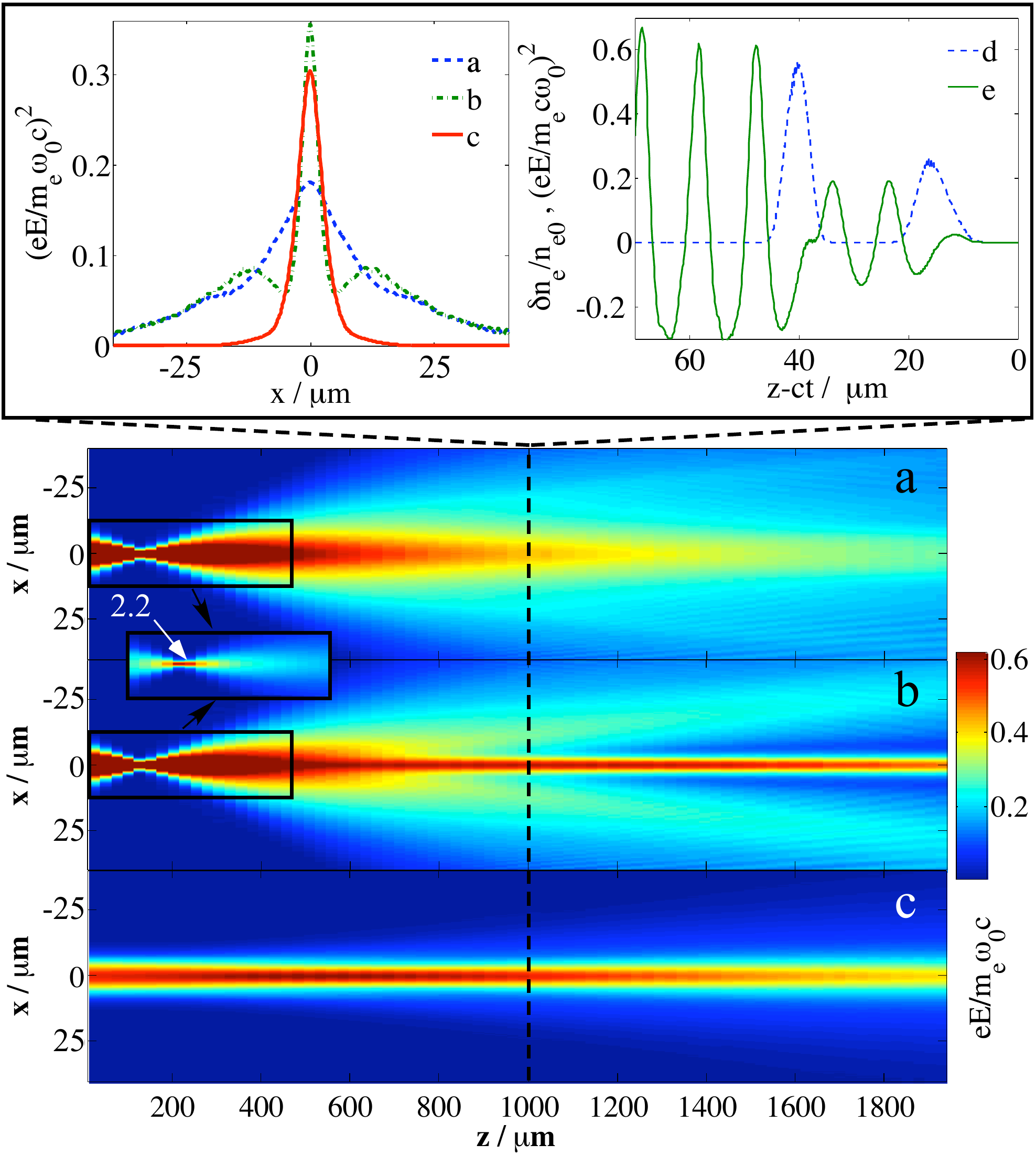}
\caption{(color online) Electric field of pulse averaged longitudinally ($z$) and then plotted as a function of propagated distance for a) driver with no guide pulse, b) driver with guide pulse and c) the guide pulse. The graphs show (left) the intensity profiles in the transverse ($x$) direction and (right) the wakefield and pulse profiles along the axis of propagation. d) is the laser envelope and e) is $\delta n_e/n_{e0}$ for the wake.}
\label{timeyhist}
\end{center}
\end{figure}
In Fig.~\ref{timeyhist}, time histories of the intensity profiles of (a) the unguided drive pulse, (b) the guided drive pulse and (c) the guide pulse are shown as they propagate through the plasma. The guided drive pulse (b) is in phase with the wake, and is clearly guided over a distance longer than its Rayleigh length ($z_R=33\;\rm{\mu m}$). The non-ideal pulse lengths (similar to those of the experiment) meant that not all of the driver is within the focusing phase of the plasma wave, and hence it diffracts away. However, the majority of the pulse energy is trapped in a single filament.

The inset graphs to Fig.~\ref{timeyhist} show the intensity profiles for the pulses and wake density profile after $1.05$~mm propagation. A gaussian fit to the temporally averaged central filaments gives a {\it full} width at $1/e^2$, $2w$ for each of the pulses. The guide pulse (c), is very close to $\lambda_p$ in spotsize, $2w=9.0\;\rm{\mu m}$, as is expected for a self-guided short pulse \cite{thomas, wei_long}. For the unguided driver pulse (a), the spotsize is $2w=20.8\;\rm{\mu m}$ and has significant wings, whereas for the guided driver pulse (b), the spotsize is $2w=5.6\;\rm{\mu m}$. 

Earlier in the interaction (a and b, inset boxes), the drive pulse focuses to high intensity and traps an electron bunch. After defocusing, electron trapping ceases, and the behavior is strongly influenced by the presence of the guide pulse. In the unguided case, the lack of quasistatic wakefields maintained for a long propagation length results in a broadening of the electron spectra c.f. \cite{thomas}. However, in the guided case, the electron bunch is accelerated in a uniform wakefield, resonantly generated by the two pulses for a relatively long time, and therefore remains monoenergetic. This means the electron trapping can be controlled through the focusing of the drive pulse.

In summary, the production of monoenergetic electrons has been demonstrated for the first time by the interaction of two collinear pulses, through plasma wave guiding and resonant plasma wave growth. Under the right conditions of laser beam overlap and timing, electron beams of high quality, both in terms of energy spread and divergence, were produced. By controlling the propagation of a tightly focused driver pulse, this scheme offers the ability to control monoenergetic electron beam production and is easy to implement. Consequently this scheme may be of interest in the development of compact sources of energetic electron beams.

This work was supported by EPSRC and Alpha-X. The authors also gratefully acknowledge the \textsc{osiris} consortium (UCLA/IST Lisboa/USC) for the use of \textsc{osiris}.

$\dag$ Present address: Ohio State University, Columbus, Ohio, US.

$\ddag$ Present address:  The University of Michigan, Ann Arbor, Michigan, US.

\end{document}